\documentclass[twocolumn,showpacs,prb,floatfix,superscriptaddress,
citeautoscript ,cite]{revtex4}
\usepackage{graphicx}
\usepackage{color}
\usepackage{amsmath}

\usepackage{booktabs}

\begin{document}

\title{Nitrogenated, Phosphorated and Arsenicated Monolayer 
Holey Graphenes}

\author{M. Yagmurcukardes} 
\email{mehmetyagmurcukardes@iyte.edu.tr}
\affiliation{Department of Physics, Izmir Institute of Technology, 35430 Izmir, 
Turkey}

\author{S. Horzum}
\affiliation{Department of Physics, University of Antwerp, 2610, Antwerp, 
Belgium}
\affiliation{Department of Engineering Physics,
Ankara University, 06100, Ankara, Turkey}

\author{E. Torun}
\affiliation{Department of Physics, University of Antwerp, 2610, Antwerp, 
Belgium}

\author{F. M. Peeters}
\affiliation{Department of Physics, University of Antwerp, 2610, Antwerp, 
Belgium}

\author{R. T. Senger}
\email{tugrulsenger@iyte.edu.tr }
\affiliation{Department of Physics, Izmir Institute of Technology, 35430 Izmir, 
Turkey}

\date{\today}


\begin{abstract}

Motivated by a recent experiment that reported the 
synthesis of a new 2D 
material nitrogenated holey graphene (C$_2$N) [Mahmood \textit{et al., Nat. 
Comm.}, 2015, \textbf{6}, 6486], electronic, magnetic, and mechanical 
properties of nitrogenated (C$_2$N), phosphorated (C$_2$P) and arsenicated 
(C$_2$As) monolayer holey graphene structures are investigated 
using first-principles calculations. Our total 
energy calculations indicate that, similar to the C$_2$N 
monolayer, the formation of the other two holey structures are also 
energetically feasible. Calculated cohesive energies for each monolayer 
show a decreasing trend going from C$_2$N to C$_2$As structure. Remarkably, 
all the holey 
monolayers are direct band gap semiconductors. Regarding the mechanical 
properties 
(in-plane stiffness and Poisson ratio), we find that C$_2$N has the highest 
in-plane stiffness and the largest 
Poisson ratio among the three monolayers. In addition, our calculations reveal 
that for the C$_2$N, 
C$_2$P 
and C$_2$As monolayers, creation of  N and P defects changes the
semiconducting behavior to a metallic ground state while the inclusion of 
double H impurities in all holey structures 
results in 
magnetic ground states. As an alternative to the 
experimentally synthesized 
C$_2$N, C$_2$P and C$_2$As are 
mechanically stable and flexible semiconductors which are important 
for potential 
applications in optoelectronics.

\end{abstract}

\maketitle

\section{Introduction}

In the last decade, graphene, one atom thick form of carbon atoms arranged in a 
honeycomb  structure, has become important in materials science 
due to its exceptional 
properties\cite{novo1,Geim1}. It has triggered 
interest in novel two 
dimensional structures such as hexagonal 
monolayer crystals of III-V binary 
compounds\cite{Novo3,hasan1} and transition metal 
dichalcogenides 
(TMDs)\cite{Novo3,Gordon,Wang}. Hexagonal monolayer structures of 
BN\cite{Zeng,Song} and 
AlN\cite{Bacaksiz,Zhuang,QWang,KKim,MFarahani} are wide 
band gap semiconductors with nonmagnetic ground states. However, monolayer 
crystals of TMDs such as MoS$_2$, WS$_2$, MoSe$_2$, WSe$_2$, MoTe$_2$ have 
direct band gaps in a favorable range of 
1-2 eV, and they are 
potentially important for optoelectronic device technology.

The lack of a band gap is 
the 
major obstacle for the use of graphene in electronic applications such 
as 
field-effect transistors\cite{XWang}, and electrodes in solar 
cells\cite{Xue,Osella}.
Thus, tuning its electrical properties through opening of a band gap is of 
great 
technological importance\cite{XWang,CBerger,Wei}. Nitrogen doping has 
been widely studied as one of the most feasible methods to 
modulate the electronic and other properties of graphene and its 
derivatives\cite{Dai1,Zhang2013,Dai2,Yu,Gong,LZhao}. 

A series of 
covalent organic frame-works (COFs)\cite{Xiang,Schlütter,Colson,Colson1,Feng} 
have been designed to form large graphene-like honeycomb networks. 
In 2005, Yaghi \textit{et al}. demonstrated the utility of the
topological design principle in the synthesis of porous organic
frameworks which are connected with covalent bonds, which are the first
successful examples of these COFs\cite{Cote}. Since COFs are composed of  
light-weight elements linked by strong covalent bonds, they have low mass 
densities and possess high thermal stability. The successful realization of  
COFs with molecular graphene-type building blocks would provide covalent 
frameworks that 
could be functionalized into light-weight materials optimized for gas 
storage, photonic, and catalytic applications\cite{Kaderi,Han}.

N-doped graphene-like honeycomb structures are important examples of COF 
materials. In a recent study by Mahmood 
\textit{et al}.\cite{Mahmood} the design and preparation of a two dimensional  
holey 
crystal, C$_2$N, 
with uniform holes and nitrogen atoms was reported. The structure and 
band gap of 
C$_2$N were studied by using both experimental techniques and 
DFT-based 
calculations. 
This new structure is layered like graphite with a different interlayer 
distance 
and is highly crystalline. It exhibits a direct band gap which was determined 
as 1.96 eV by using ultraviolet visible 
spectroscopy, while a 
slightly 
smaller band gap of 1.70 eV is 
obtained from density functional theory (DFT) calculations. In another study, 
Sahin 
investigated 
the structural and phononic characteristics 
of the 
C$_2$N structure\cite{H.Sahin}. The formation of heterostructures of holey 
graphenes 
and the resulting Moir\'{e} patterns were investigated by Kang \textit{et 
al}\cite{Jun}. Very recently Zhang \textit{et 
al.} investigated the structural and electronic properties of few-layer C$_2$N 
by considering different stacking orders and number of layers\cite{zhangrev}. 
In the study by Xu \textit{et al.} energy barriers for the adsorption of H$_2$, 
CO$_2$ and CO molecules on C$_2$N monolayer were calculated for a possible 
H$_2$ dissociation\cite{xu2015}.

Motivated by the recent experiment on synthesis of C$_2$N 
monolayer\cite{Mahmood} and by the studies on graphene-like networks composed 
of COFs, we investigate the structural, electronic and mechanical 
properties of 2D holey crystals of C$_2$X (X=N, P or As) stoichiometry. The 
mechanical properties of these hexagonal structures are examined under uniaxial 
strain, and the in-plane stiffness and the Poisson ratio values are obtained. 
In addition, the most probable types of atomic scale disorder, formation of N, 
P and As 
defects, are investigated for these holey structures. 

The paper is organized as follows: Details of the computational methodology are 
given in Sec. 
\ref{comp}. Structural properties of C$_2$N, C$_2$P and C$_2$As are 
presented in Sec. 
\ref{structural}. Discussions about electronic and magnetic properties of 
these monolayer crystals are given in Sec. \ref{electronic}. In Sec. 
\ref{mechanical} the mechanical properties are discussed by examining the 
in-plane stiffness and the Poisson ratio for each structure. Electronic and 
geometric properties of 
defect and H-impurities in C$_2$X monolayers are discussed in Sec. 
\ref{defectsec}. Finally we 
conclude in Sec. \ref{Conc} 
 
\begin{figure}[htbp]
\centering
  \includegraphics[width=0.4\textwidth]{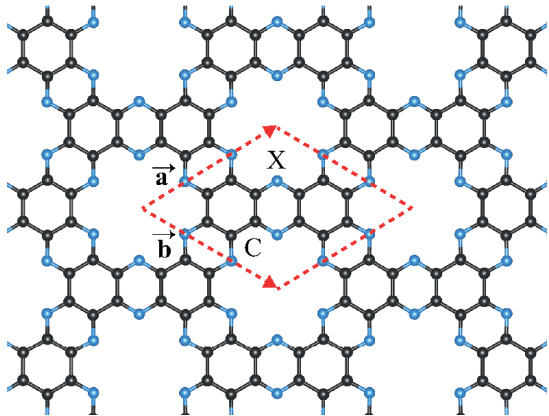}
  \caption{Top view of C$_2$X holey graphene monolayer structure where X 
represents N, P or As atoms.}
  \label{stuc}
\end{figure}

\begin{table*}[htbp]
\small
  \caption{\label{table} The calculated ground state properties of 
C$_2$X-structures, structural geometry, 
 lattice 
parameters of primitive unit cell, $a$ (see Fig. \ref{stuc}); the 
distance between C-X 
atoms, $d_{C-X}$; the distance between two 
carbon atoms, $d_{C-C}$; 
angle for C-X-C bond, $\theta_{CXC}$; magnetic state; the average charge 
donated 
to ($+$) of from ($-$) 
each C atom, $\Delta\rho$; the cohesive 
energy per atom in primitive unitcell, $E_{\rm coh}$; the 
energy band gap of the 
structure calculated within, GGA with the inclusion of SOC, $E_{\rm g}^{GGA}$; 
and 
HSE06, $E_{\rm g}^{HSE}$;
workfunction, $\Phi$; Poisson's ratio, $\nu$; and in-plane 
stiffness 
$C$. Calculated parameters for graphene and h-BN are given for comparison.}
\begin{tabular*}{1.0\textwidth}{rcccccccccccccccc}
\hline
& $a$ & $d_{\rm {C-X}}$ & $d_{\rm {C-C}}$ & $\theta_{\rm CXC}$ 
& Magnetic & 
$\Delta\rho$ 
& $E_{\rm coh}$ & $E_{\rm g}^{GGA}$ &  $E_{\rm g}^{HSE}$ & 
$\Phi$ & $\nu$ & $C$ &\\
& (\AA{}{})  & (\AA{}{}) & (\AA{}{}) & (deg) & State & 
($e$) & (eV) & (eV) & (eV) & 
(eV) &  
&(eV/\AA{}$^2$)&\\
\hline
C$_2$N &  8.33 &  1.34 & 1.47 & 118 & NM & $-$0.6 & 7.64 & 
1.66(d) &  2.47(d) & 5.23 & 0.26 & 9.27 \\
C$_2$P &  9.33 &  1.76 & 1.42 & 108  & NM & $+$0.6 & 6.84 & 
0.22(i) & 0.94(d) & 4.90 & 0.21 & 6.69 &\\
C$_2$As &  9.72 &  1.92  & 1.41 & 108  & AFM & $+$0.3 & 5.78 
&  0.43(d) & 1.16(d) & 4.89 & 0.21 & 5.83 &\\
Graphene &  2.46 & - & 1.42 & 120 & NM & 0.0 & 7.97 & 
- & - & 4.51 & 0.16\cite{CLee} & 21.25\cite{CLee} \\
h-BN &  2.51 & 1.45 (B-N) & - & 120 & NM & $+$2.1 & 7.10 & 
4.48(d) & 5.56(d)\cite{Berseneva} & 5.80 & 0.22 & 17.12\\
\hline
\end{tabular*}
\end{table*}

\section{COMPUTATIONAL METHODOLOGY}\label{comp}
First-principles calculations were performed within the 
framework of density functional theory (DFT) by using the Vienna Ab initio 
Simulation Package (VASP)\cite{vasp1,vasp2,vasp3,vasp4}. The  approach 
is based on an iterative solution of the Kohn-Sham equations\cite{Kohn-Sham} 
with a plane-wave set adopted with the Perdew-Burke-Ernzerhof (PBE) 
exchange-correlation functional of the generalized gradient approximation (GGA) 
\cite{GGA-PBE1,GGA-PBE2} with the inclusion of spin-orbit-coupling (SOC). More 
accurate results, electronic structure
calculations were performed using the Heyd-Scuseria-Ernzerhof (HSE) 
screened-nonlocal-exchange functional of the generalized Kohn-Sham scheme
\cite{HSE}. Analysis of the charge transfers in the 
structures was made by the Bader 
technique\cite{Henkelman}. 

Electronic and geometric relaxations of hexagonal monolayers of C$_2$X 
structures
were performed by considering the following criteria in our 
calculations. The energy cut-off value for the plane wave basis set was taken 
to be 
$500$ eV. The energy difference between sequential steps for the electronic 
self 
consistence-loop 
was 
considered to be $10^{-5}$ eV. As a convergence criterion in the structural 
relaxation and for the
Hellmann-Feynman forces on each atom was taken to be 0.05 eV/\AA {}. For 
geometric relaxation of the structures 
a parallelogram unit cell containing 12 C atoms and 6 X atoms was used (see 
Fig. \ref{stuc}). The minimum energy was 
obtained by varying the lattice constant and the pressure was reduced below 1 
kbar. Brillouin zone 
integration was performed by using a set of $5\times5\times1$ $\Gamma$-centered 
k-point sampling mesh for a single unit cell. To get more accurate results 
for the density of states (DOS) and the work 
function calculations a 
set of $15\times15\times1$ k-point sampling was used. The broadening for DOS 
calculations was taken to be 
0.05. The cohesive energy per 
atom in a primitive unit cell was calculated 
using the formula;
\begin{equation}
\label{equ}
 E_{\rm coh}=[12E_{\rm C}+6E_{\rm X}-E_{\rm C_{2}X}]/18
\end{equation}
where $E_{\rm C}$ and $E_{\rm X}$ 
denote the 
magnetic ground state energies 
of the single C and X atoms, respectively while $E_{\rm C_{2}X}$ denotes the 
total energy of the monolayer C$_{2}$X. Calculations on elastic 
constants 
were performed by considering 
a $2\times2$ supercell containing 72 atoms.


\section{STRUCTURAL PROPERTIES}\label{structural} 
Generic forms of the monolayer structures, C$_2$N, C$_2$P and 
C$_2$As, display honeycomb 
symmetry as shown in Fig. \ref{stuc}. All calculated 
parameters for their relaxed geometries are listed in 
Table \ref{table}.
In the primitive unit cell there are 12 C atoms and 6 X atoms, X being N, P or 
As. The C$_2$N crystal has a planar two dimensional 
structure with a lattice constant of 8.33 \AA{} {} which is consistent 
with the value reported by Mahmood \textit{et al}.\cite{Mahmood} The 
calculated C-C bond length is 1.47 \AA{} {} while 
the C-N bonds are 1.34 \AA{} {} with the C-N-C bond angle being 118 degrees. 
This 
bond 
angle is the largest one of all three structures. 
This means that the hole between the benzene rings is nearly a perfect 
hexagon in C$_2$N. Bader charge
analysis shows that an average 0.6 $e$ of charge depletion per atom occurs 
from 
C atoms to the neighboring N atoms. The cohesive energy per atom is 
highest for the C$_2$N 
structure with a value of 7.64 eV, as calculated using Eq. (\ref{equ}). 
 
Optimized lattice constant of the C$_2$P monolayer crystal is calculated to 
be 9.33 \AA{}. The C-C bond 
length is 1.42 \AA{} as in graphene hexagons and the C-P bond length is 1.76 
\AA{} {}. The C-P-C bonds have a narrower angle than that of the 
C-N-C bonds with a value of 108 degrees. According the Bader 
charge 
analysis, opposite to the C$_2$N case an average of 0.6 $e$ charge is 
transfered to each C atom from the P atoms.
The cohesive energy per atom, 6.84, eV is less than that of 
C$_2$N.
\begin{figure}[!htbp]
\includegraphics[width=0.4\textwidth]{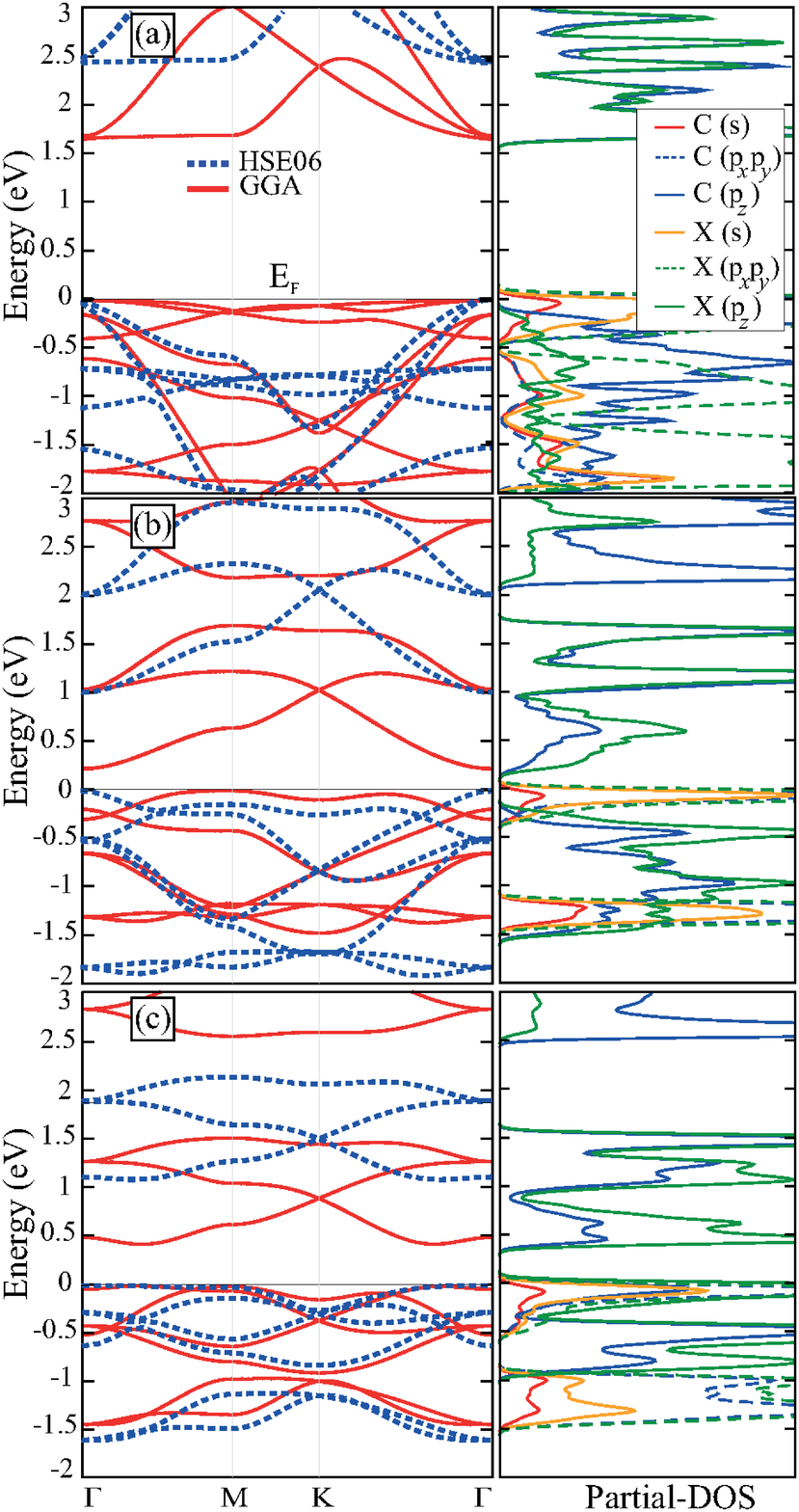}
\caption{\label{bands} Band-structures (left-panel) and corresponding partial 
density 
of states (right-panel)
of (a) C$_2$N (b) C$_2$P, and (c) 
C$_2$As where red curves are for bands calculated within GGA approximation 
while 
dashed blue curves are for bands calculated within HSE06 on top of GGA. The 
Fermi energy ($E_F$) 
level 
is set to the valence band maximum.}
\end{figure}

For the C$_2$As monolayer structure the lattice constant is 9.72 
\AA{} with a corresponding C-C 
bond length of 1.41 \AA{} which is nearly the same as the C-C bond in C$_2$P. 
The 
longest bond 
length between a C atom and its X neighbor is found for the C-As bond with a 
value of 1.92 \AA{}. The angle between 
two 
C-As bonds, 108 degrees, is smaller than that of the C-N bonds. 
We found that the charge 
transfer occurs in this structure from As atoms to each C atom with a value of 
0.3 $e$.
The charge transfer occurs from P and As 
atoms to the C rings for C$_2$P and C$_2$As, respectively. However, it occurs 
from C rings to the N 
atoms in C$_2$N. 

\section {ELECTRONIC AND MAGNETIC PROPERTIES}\label{electronic}

The calculated lattice constant and electronic band gap of C$_2$N 
are in agreement with the values reported by Mahmood \textit{et. 
al}.\cite{Mahmood}
C$_2$N monolayer has a 
direct band gap of 1.66 eV and 2.47 eV in GGA and HSE06 levels, respectively as 
seen in Fig. 
\ref{bands}(a). The overall dispersion characteristic of the bands are not 
affected by the inclusion of HSE06.
The valence band maximum (VBM) and 
conduction band minimum (CBM) of the C$_2$N 
monolayer lies at the $\Gamma$ point of the Brillouin 
zone. Relatively large value of the C$_2$N energy band gap makes it a suitable 
semiconductor for various device applications. It also appears from the energy 
band structure that spin up and spin down states are degenerate 
throughout the 
Brillouin Zone and therefore the structure does not exhibit any spin 
polarization in its ground
state. Due to the pairing of \textit{p}$_z$ electrons of 
3-coordinated C atoms and 2-coordinated N atoms, the structure has a 
nonmagnetic ground state.

The electronic band dispersion for the C$_2$P monolayer crystal indicates that 
it has an indirect band gap of 0.22 eV and a direct gap of 0.94 eV in GGA and HSE06 levels, 
respectively. 
Since the VBM of C$_2$P monolayer consists of localized states, these states 
are 
affected by the HSE06 functional and the VBM 
point of the band structure moves to the $\Gamma$ point as 
shown in Fig. \ref{bands}(b). Like the C$_2$N monolayer, C$_2$P has 
also 
a nonmagnetic ground state.


In Table I, the workfunctions of the monolayer holey 
graphenes are also shown. It is seen that the workfunction values of these 
compounds are smaller than that of the h-BN and larger than the value for 
graphene. Comparing the 
workfunction values of the monolayers a decreasing trend 
can be seen from nitrogenated one to the arsenicated one. This result can be 
explained 
by the decreasing ionization energy of the elements in the periodic table from 
top to bottom rows.

\begin{figure}[htbp]
\centering
\includegraphics[width=0.4\textwidth]{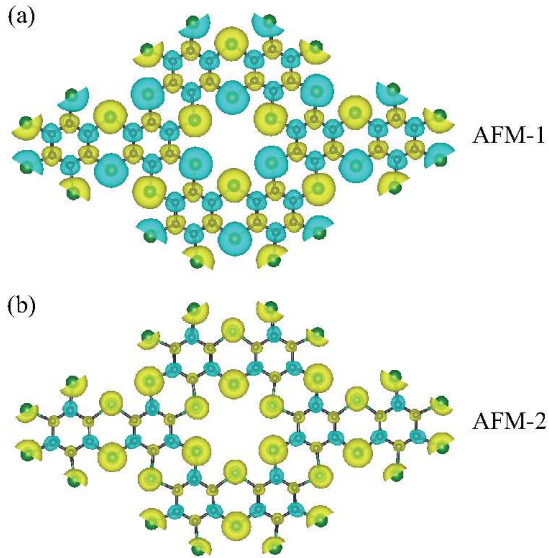}
\caption{\label{dens}
(Color online) Charge density difference, $\rho_{up}$-$\rho_{down}$, of C$_2$As 
monolayer structure for (a) 
fully anti-ferromagnetic (AFM-1) order and (b) anti-ferromagnetic order in the 
benzene ring (AFM-2) where green/yellow color is for minority/majority 
spin states. The plotted isosurface 
values are 10$^{-3}$ e/\AA{}$^3$ and 10$^{-5}$ e/\AA{}$^3$ for (a) and (b) 
respectively.}
\end{figure}

The C$_2$As monolayer crystal is a semiconductor with a direct band gap of 0.43 
eV and 1.16 eV in GGA and HSE06 levels, respectively. Similiar to the case of 
C$_2$N, the inclusion of 
HSE06 functional increases the energy gap and does not change the dispersion 
characteristic of the band structure (see Fig. 
\ref{bands}(c)). 
Both the VBM and the CBM of C$_2$As lie between the $\Gamma$ and the M 
points. Interestingly, there is an isolated Dirac point in the 
conduction band of C$_2$As which can be populated using doping or a gate 
potential. 
The net magnetic moment for this structure 
is zero like for the other two 
monolayers. But the ground state is obtained for anti-ferromagnetic 
(AFM-1) ordering given 
in 
Fig. \ref{dens}(a) in which all the neighboring C and As atoms have equal but 
opposite local 
magnetic moments in their sublattices. In the AFM-2 magnetic ordering, 
the C atoms in a ring, have opposite magnetic moments while the As atoms have 
ferromagnetically ordered moments
as seen in Fig. \ref{dens}(b). The net magnetic moments of the two 
configurations, AFM-1 and AFM-2, are zero with an energy difference of 50 
meV, AFM-1 being the ground state.

\begin{figure*} [htbp]
\centering
\includegraphics[width=0.95\textwidth]{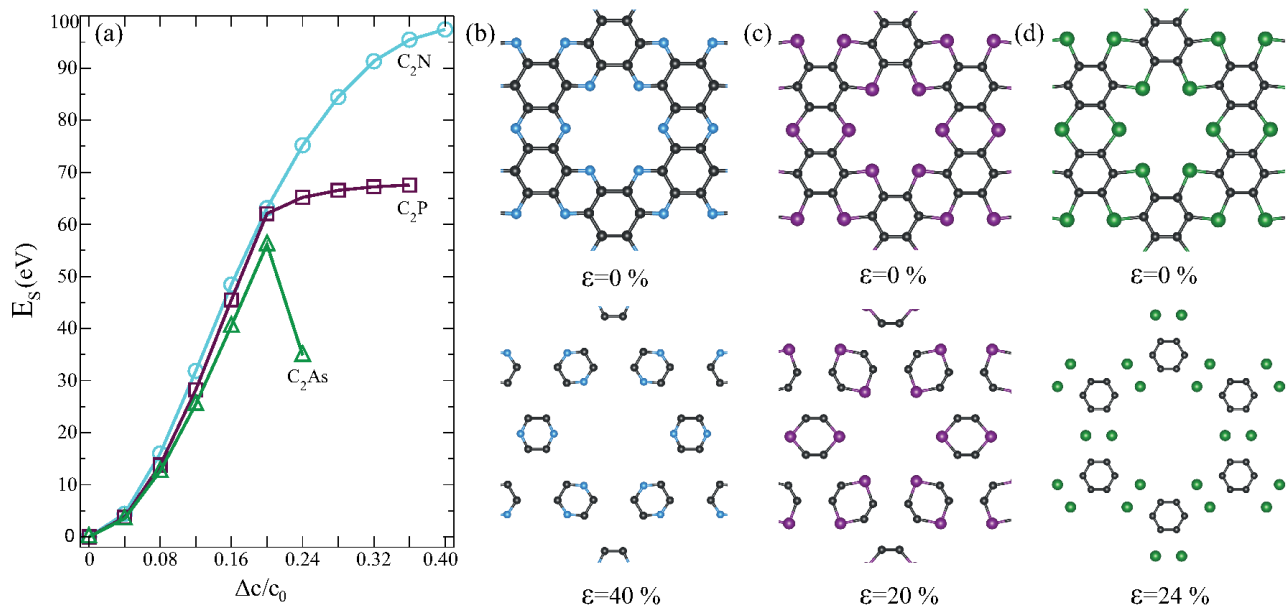}
\caption{\label{mech}
{(a) Change of total energy of the three holey graphene 
monolayers under applied 
strain. Structural changes under applied strain of (b) C$_2$N, 
(c) C$_2$P and (d) C$_2$As.} }
\end{figure*}

\section{MECHANICAL PROPERTIES}\label{mechanical}

The elastic properties of homogeneous and isotropic materials can be 
represented by two independent constants, 
the in-plane stiffness $C$ and the Poisson ratio $\nu$. The stiffness 
parameter is a measure of the rigidity or the flexibility of a material. The 
mechanical 
response of a material to an applied 
stress is called the Poisson ratio. It is also defined as the ratio 
of the transverse contraction strain to the longitudinal extension strain in 
the 
direction of the stretching force, that is 
$\nu$$=$$-$$\varepsilon$$_{trans}$/$\varepsilon$$_{axial}$. 

To calculate the elastic constants of 
C$_2$N, C$_2$P and C$_2$As monolayers, a $2\times2$ 
supercell containing 72 atoms is considered. 
The strains ${\varepsilon}_x$ and 
${\varepsilon}_y$ are applied to the monolayer crystals by varying the lattice 
constants 
along the x and y directions. The strain parameters ${\varepsilon}_x$ and 
${\varepsilon}_y$ are varied between $\pm0.02$ with a step size of 
0.01. For this purpose three different sets of data are calculated; 
(i) ${\varepsilon}_y$=0 and ${\varepsilon}_x$ varying, 
(ii) ${\varepsilon}_x$=0 and ${\varepsilon}_y$ varying and (iii) 
 ${\varepsilon}_x$=${\varepsilon}_y$. At each configuration, 
the atomic positions 
are fully relaxed  and the strain energy, $E_S$, is calculated by subtracting 
the total energy 
of the strained system from the equilibrium  total  energy. The calculated data 
is fitted to the equation 
$E_S=c_1{{\varepsilon}_x}^2+c_2{{\varepsilon}_y}^2+c_3{\varepsilon}_x{ 
\varepsilon}_y$, so that the coefficients c$_i$ are determined. The in-plane 
stiffness $C$ can then 
be calculated from $C=(1/A_0)(2c-{c_3}^2/2c)$ where we let c$_1$$=$c$_2$$=$c 
due 
to isotropy of the unit cell and A$_0$ is the 
unstretched 
area of the $2\times2$ 
supercell. The Poisson ratio is obtained as $\nu=c_3/2c$. Due to the 
symmetry of the 
honeycomb lattice, the in-plane stiffness and the Poisson ratio are the same 
along the x and y directions.

As indicated in Table \ref{table}, the calculated in-plane stiffness for C$_2$N 
is 
9.27 eV/\AA{}$^2$ which has the highest value 
among the three monolayer structures. This value indicates a 
strong bonding between the C and N atoms. Although it is the highest value, 
it is still smaller compared to that of graphene and h-BN\cite{Clee}. The 
calculated Poisson ratio for the C$_2$N monolayer is 0.26 which is in the range 
for usual two dimensional materials. This means that when the material is 
compressed in one direction, it will expand in the other direction as well. The 
in-plane stiffness value is for C$_2$P is calculated (6.69 
eV/\AA{}$^2$) with a corresponding Poisson ratio of 0.21, which means 
that the C$_2$P crystal is less responsive than C$_2$N under compression. The 
lowest in-plane stiffness for the C$_2$As monolayer is 5.83 eV/\AA{}$^2$ with 
the corresponding Poisson ratio of 0.21 which is equal 
to that of the C$_2$P crystal. All the holey monolayers have Poisson ratios 
which are larger than that of graphene and close to that of h-BN.

We next consider the behavior of the monolayer structures under higher values 
of uniform strain ranging from 0.04 to 0.40. For this purpose 
the calculations are performed in a $2\times2$ supercell. The 
change of 
strain energy of all the monolayers under applied 
biaxial 
strain is given in Fig. 
\ref{mech}(a).
Although C$_2$N is the stiffest crystal, structural 
deformations start to form beyond 
12\% strain which is small compared to those of C$_2$P and C$_2$As. By 
structural 
deformation we mean that 
the N atoms connecting 
the C pairs start to form C$_4$N$_2$ isolated hexagonal rings. The distance 
between two 
neighboring 
C atoms in different hexagonal rings become 1.79 \AA{} {}
at 12\% strain. This distance increases up to 
3.10 \AA{} {} at 40\% strain. The 
deformation path seems to be the same for the C$_2$P monolayer structure. Up to 
a strain 
value of 20\%, P atoms are 
still 
bonded to the hexagonal C rings and there is no drastic change in the structure 
of the 
monolayer. However beyond 20\% strain hexagonal rings are formed 
composed of 
4-C 
and 2-P atoms as in the case of 
C$_2$N. The C-C bond lengths in C pairs are approximately 1.30 \AA{} {} at 
20\% 
strain and there exist C-C pairs connected by P atoms as shown in 
Fig. \ref{mech}(c). 
Among the monolayer structures considered, only in the C$_2$As crystal the 
hexagonal C rings 
preserve their form under large strains. The 
bond angle 
of C-As-C bond gets larger as the applied strain is increased. As 
given 
in Fig. \ref{mech}(d), at 24\% 
strain this angle becomes 
158 degrees and there is no longer bonding between the C and As atoms. 
Compared with the other two 
structures, C$_2$As has the smallest in-plane stiffness value and it is the 
softest material among the three monolayers. The C$_2$X monolayer structures 
can 
be viewed as an ordered phase of 6-C rings 
linked by the X atoms. It seems that the linker atoms N and P have stronger 
bonds to their C neighbors so that the structure 
dissociates into isolated rings by breaking the C-C bonds under high strain. 
For 
As, however, the C-C bonds must be
stronger than the C-As bonds so that the crystal yields at the linker sites. 

\begin{figure}[htbp]
\centering
\includegraphics[width=0.45\textwidth]{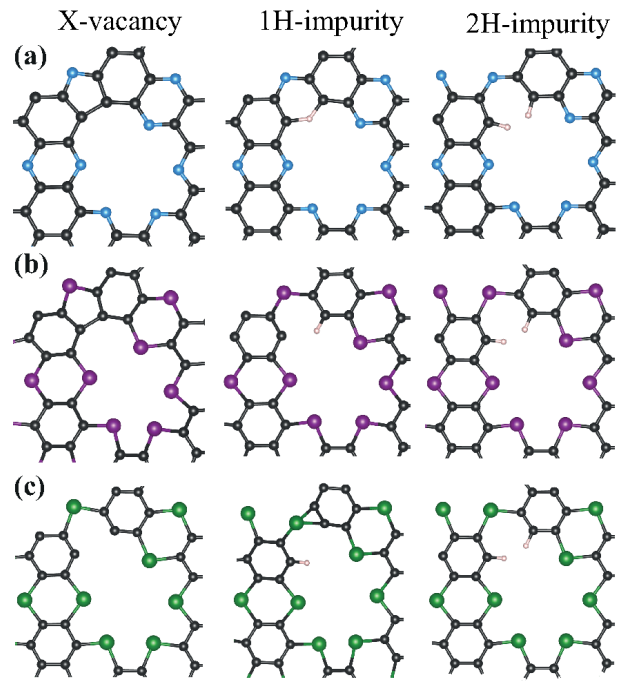}
\caption{\label{defect}
(Color online) Optimized X-vacancy and their H substituted structures of (a) 
C$_2$N, 
(b) C$_2$P, (c) C$_2$As respectively.}
\end{figure}

\section{EFFECT OF DEFECTS}\label{defectsec}
Considering the synthesis procedure of the mentioned holey crystals in which 
the ingradient molecules are self-assembled, 
the atomic scale disorders like vacant N, P and 
As sites are the most probable disorders in C$_2$N, C$_2$P and 
C$_2$As monolayers, 
respectively. The existence of H-impurities at these 
vacant-sites 
are also possible since the C$_2$N holey structure is  synthesized as a result 
of the interactions of 
hexaaminobenzene and 
hexaketocyclohexane molecules which contain H atoms in their composition. In 
this section, we investigate the effects of 
these 
vacant sites and substitutional H-impurities on the 
geometric and the electronic properties of the monolayer holey structures.

\begin{figure}[htbp]
\centering
\includegraphics[width=0.45\textwidth]{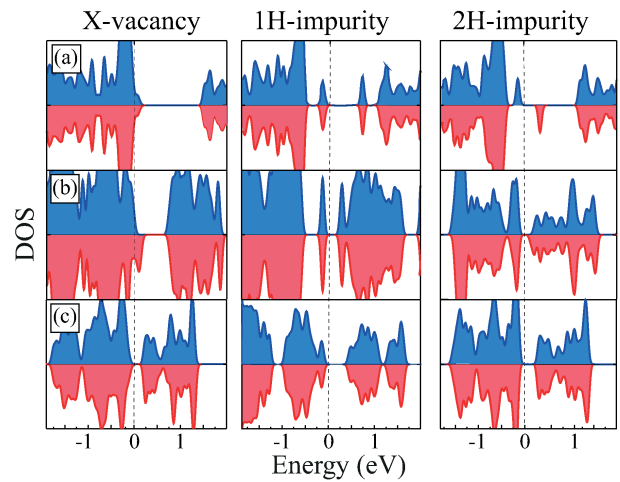}
\caption{\label{imp}
(Color online) Total DOS for defected and H-impurity structures of (a) C$_2$N, 
(b) C$_2$P, and (c) C$_2$As respectively. }
\end{figure}

Optimized geometries of the defected structures 
are shown in 
Fig. \ref{defect}. For the N-defected C$_2$N and P-defected C$_2$P holey 
crystals (Figs. \ref{defect}(a) and (b)), 
removal of a single N or P atom results in a bond formation between the two C 
atoms at the vacant site. However, in the case of As-defected C$_2$As the 
optimized geometric structure 
does not lead to an additional bonding (see Fig. 
\ref{defect}(c)). 
Geometry optimizations indicate that for X-vacant structures only 
the C$_2$N retainsits planar geometry while the other two structures get 
buckled. 
Our Bader analysis shows that charge depletion of 1.1 $e$ per atom occurs from 
the C atoms to the neighboring N atoms in N-defected C$_2$N.
For P-defected C$_2$P, an average of 0.6 $e$ charge is transferred to each C 
atom 
except 
for the two C atoms at the vacant sites since these C atoms keep approximately 
their initial charges. For As-defected 
C$_2$As the charge is depleted to the C atoms with a value of 0.3 $e$ per 
atom on the average.
We found that the N-defected C$_2$N has a nonmagnetic ground state 
while P-defected C$_2$P and As-defected 
C$_2$As have magnetic ground states with a net moment of 1 $\mu_B$. Total DOS 
calculations indicate that X-missing structures of 
C$_2$N and C$_2$P become metallic monolayers while the As-defected C$_2$As is 
still a semiconductor with a lower band gap energy than its 
perfect form (see Fig. \ref{imp}(c)). Calculated cohesive energies per atom in 
the 
supercells demonstrate that for all three structures 
the highest $E_{\rm coh}$ occur for X-missing structures of C$_2$N and 
C$_2$P while for the C$_2$As monolayer most energetic case 
is 1H-impurity case as seen in Table \ref{table2}.

\begin{table}[htbp]
\caption{\label{table2} The calculated ground state properties of defected 
C$_2$X-structures, structural geometry, 
 lattice 
parameters of $2\times2$ supercell, $a$ and $b$, the net magnetic moment 
of the structure, $\mu$, and the cohesive 
energy per atom in supercell $E_{\rm coh}$.}
\begin{tabular}{rcccccccccccccccc}
\hline\hline
& Geometry & $a$ & $b$ & $\mu$ & $E_{\rm coh}$ \\
& & (\AA {}) & (\AA {}) & ($\mu_{B}$) & (eV) \\
\hline
N-vacant-C$_2$N & planar & 16.44 & 16.44 & 0 & 6.77 \\
P-vacant-C$_2$P & buckled & 17.66 & 17.67 & 1 & 6.02 \\
As-vacant-C$_2$As & buckled & 19.27 & 19.27 & 1 & 5.61 \\
\hline
1H-imp.-C$_2$N & planar & 16.63 & 16.63 & 0 & 6.71 \\
1H-imp.-C$_2$P & buckled & 18.37 & 18.15 & 0 & 5.97 \\
1H-imp.-C$_2$As & buckled & 18.37 & 19.04 & 0 & 5.65 \\
\hline
2H-imp.-C$_2$N & planar & 16.79 & 16.79 & 1 & 6.67 \\
2H-imp.-C$_2$P & planar & 18.71 & 18.71 & 1 & 5.93 \\
2H-imp.-C$_2$As & planar & 19.46 & 19.45 & 1 & 5.58 \\
\hline\hline
\end{tabular}
\end{table}

As seen in Fig. \ref{defect}(a), the C$_2$N structure having a single H 
substitution 
at the 
N-vacant site preserves the geometry of C$_2$N monolayer. The geometries of 
other two monolayers with 1H-impurity get buckled as seen in Fig. \ref{defect}.
The result of the Bader analysis for all three monolayers for 1H-impurity case 
show 
that 0.1 $e$ of charge is transferred to the C atom at the vacant site from H 
atoms. The inclusion of single 
H-impurity gives rise to a non-magnetic 
ground state for all C$_2$X monolayers as 
in their bare cases. For 1H-impurity structures the total DOS calculations 
indicate that all three monolayers 
preserve their semiconducting character but with lower values of band gap 
energies (see Fig. \ref{imp}).

In our study, inclusion of 2H-impurities at the X-vacant sites 
is also considered. 
In all three defected structures each H atom binds to a single C atom as 
expected (see Fig. \ref{defect}). Addition of the second H atom to the vacancy 
sites restores the planar geometry 
of all three monolayers. In this case, the charge is donated 
to each C atom at the vacant site from the H atoms such that the final charges 
on C atoms are the same 
as their values in the perfect crystals. 
2H impurities 
result in a magnetic ground state for 
all three monolayers with a 1 $\mu_B$ of net magnetic moment. The total DOS 
calculations demonstrate that 
inclusion 2H-impurities in C$_2$X structures preserves the semiconducting 
behaviors of the three monolayers with lower band gap energies (see Fig. 
\ref{imp}). The corresponding cohesive energies per atom are also
given in Table 
\ref{table2} which indicate that the C$_2$N monolayer has the highest $E_{\rm 
coh}$ than that of C$_2$P and C$_2$As 
monolayers.

\section{CONCLUSIONS}\label{Conc}
Motivated by recent experiments on the C$_2$N monolayer and 
graphene-like COF networks, we investigated structural, mechanical and 
electronic properties of two other monolayer structures, C$_2$P and 
C$_2$As. We found that C$_2$N has the highest $E_{\rm coh}$ 
among the three monolayers and the calculated values of $E_{\rm coh}$ are 
comparable 
with that of graphene and h-BN. Moreover, it is calculated that, the 
workfunction values of the monolayers are decreasing from C$_2$N to C$_2$As 
which is consistent with the trend in ionization energy of each element. 
Energy-band structure calculations show that the holey monolayers 
are direct band gap semiconductors. 
Our calculations on mechanical constants suggest that 
the stiffest material is the C$_2$N structure with the highest Poisson ratio 
among the three monolayers. Moreover, the vacancy defects of N and P atoms in 
holey structures
lead to metallic ground states
while the substitutional H-impurities do not change their semiconducting 
character but 
can create net magnetization on the monolayer. Finally, we point out that holey 
graphene monolayers are new two dimensional materials that are mechanically 
stable 
and they are flexible semiconductors which may be favorable for 
applications in optoelectronics.

\begin{acknowledgments}

This work was supported by the Flemish Science Foundation 
(FWO-Vl) and the 
Methusalem foundation of the Flemish government. Computational resources were 
provided by TUBITAK ULAKBIM, High Performance and Grid Computing Center 
(TR-Grid 
e-Infrastructure).  

\end{acknowledgments}


\begin{thebibliography}{99}


-----------------------------------------

\bibitem{novo1} K. S. Novoselov, A. K. Geim, S. V. Morozov, D. Jiang, Y. Zhang, 
S. V.
Dubonos, I. V. Grigorieva, and A. A. Firsov,
Science \textbf{306}, 666 (2004).

\bibitem{Geim1} K. S. Novoselov, A. K. Geim, S. V. Morozov, D. Jiang, M. I. 
Katsnelson,
I. V. Grigorieva, S. V. Dubonos, and A. A. Firsov,
Nature \textbf{438}, 197 (2005).

\bibitem{Novo3} K. S. Novoselov, D. Jiang, F. Schedin, T. Booth, V. V. Khot-
kevich, S. Morozov, and A. K. Geim, Proc. Natl. Acad. Science
U.S.A. \textbf{102}, 10451 (2005).

\bibitem{hasan1} H. Sahin, S. Changirov, M. Topsakal, E. Bekaroglu, E. Akturk,
R. T. Senger, and S. Ciraci, Phys. Rev. B \textbf{80}, 155453 (2009).

\bibitem{Gordon} R. A. Gordon, D. Yang, E. D. Crozier, D. T. Jiang, and R. F. 
Frindt, 
Phys. Rev. B \textbf{65}, 125407 (2002).

\bibitem{Wang}  Q. H. Wang, K. K. Zadeh, A. Kis, J. N. Coleman and M. S. 
Strano, 
Nat. Nanotechnol. \textbf{699}, 699 (2012).

\bibitem{Zeng} H. Zeng, H. Zhi, C. Zhang, Z. Wei, X. Wang, X. Guo, W. Bando, Y. 
Golberg, D. Nano Lett. \textbf{10}, 5049
(2010).

\bibitem{Song} L. Song, L. Ci, L. Lu, H. Sorokin, P. B. Jin, C. Ni, J. 
Kvashnin, 
A. G. Kvashnin, D. G. Lou, J. Yakobson, B. I. Ajayan, P. M. Nano Lett. 
\textbf{10}, 3209
(2010).

\bibitem{Bacaksiz} C. Bacaksiz, H. Sahin, H. D. Ozaydin, S. Horzum, R. T. 
Senger, and F. M. Peeters, Phys. Rev. B \textbf{91} 085430
(2015). 

\bibitem{Zhuang} H. L. Zhuang and R. G. Hennig, Appl. Phys. Lett. \textbf{101}, 
153109 
(2012).

\bibitem{QWang} Q. Wang, Q. Sun, P. Jena, and Y. Kawazoe, ACS Nano \textbf{3}, 
621
(2009).

\bibitem{KKim} K. K. Kim, A. Hsu, X. Jia, S. M. Kim, Y. Shi, M. Hofmann, D. 
Nezich, J. F. Rodriguez-Nieva,
M. Dresselhaus, T. Palacios, and J. Kong, Nano Lett. \textbf{12}, 161
(2012).

\bibitem{MFarahani} M. Farahani, T. S. Ahmadi, and A. Seif, J. Mol. 
Struct. \textbf{913}, 126
(2009).

\bibitem{XWang} X. Wang, X. Li, L. Zhang, Y. Yoon, P. K. Webe, H. Wang, J. Guo, 
Ho. Dai, 
Science \textbf{324}, 768 (2009).

\bibitem{Xue} Y. Xue, J. Liu, H. Chen, R. Wang, D. Li, J. Qu, and L. Dai, Ang. 
Chem. Int. Ed. \textbf{51}, 12124 (2012).

    
\bibitem{Osella} S. Osella, A. Narita, M. G. Schwab, Y. Hernandez, X. Feng, K. 
Müllen, D. Beljonne, ACS Nano 
\textbf{6} 5539 (2012).

\bibitem{CBerger} C. Berger, Z. Song, X. Li, X. Wu, N. 
Brown, C. Naud, D. Mayou, T. Li, J. Hass, A. N. Marchenkov,
E. H. Conrad, P. N. First, W. A. de Heer, Science \textbf{312}, 
1191 (2006).


\bibitem{Wei} D. Wei, Y. Liu, Y. Wang, H. Zhang, L. Huang, and G.Yu, Nano Lett. 
\textbf{9}, 1752 (2009).


\bibitem{Dai1} L. Dai, D. W. Chang, J. B. Baek, and W. Lu, Small \textbf{8}, 
1130 (2012). 

\bibitem{Zhang2013} J. Zhang, F. Zhao, Z. Zhang, N. Chen, and L. Qu, Nanoscale 
\textbf{5}, 3112 
(2013).


\bibitem{Dai2} L. Dai, Acc. of Chem. Res. \textbf{46} 31
(2012). 


\bibitem{Yu} D. Yu and L. Dai, J. of Phys. Chem. Lett. \textbf{1}, 467 
(2010).


\bibitem{Gong} K. Gong, F. Du, Z. Xia, M. Durstock, and L. Dai, Science 
\textbf{323}, 760 
(2009).

\bibitem{LZhao} L. Zhao et al., Science \textbf{333}, 999 (2011).

\bibitem{Xiang} Z. Xiang and D. Cao, J. Mater. Chem. A  
\textbf{1} 2691 (2003). 

\bibitem{Schlütter} F. Schlütter, F. Rossel, M. Kivala, 
V. Enkelmann, J. Pa. Gisselbrecht, P. Ruffieux, R. Fasel, 
and K. Müllen, J. of the Am. Chem. Soc.  \textbf{135}, 4550 (2013).

\bibitem{Colson} J. W. Colson, W. R. Dichtel, Nat. Chem. \textbf{5}, 453 (2013).

\bibitem{Colson1} 
    J. W. Colson,
    A. R. Woll,
    Ar. Mukherjee,
    M. P. Levendorf,
    E. L. Spitler,
    V. B. Shields,
    M. G. Spencer,
    J. Park,
    W. R. Dichtel, Science  
\textbf{332} 228 (2011). 

\bibitem{Feng} X. Feng, X. Ding, and D. Jiang, Chem. Soc. Rev.  \textbf{41}, 
6010 (2012).

\bibitem{Cote} A. P. Cote, A. I. Benin, N. W. Ockwig, M. O'Keeffe, A. J. 
Matzger, and O. M. Yaghi, 
Science \textbf{310}, 1166 (2005).

\bibitem{Kaderi} H. M. El-Kaderi, J. R. Hunt, J. L. M. Cortés, A. P. Cote, R. E. 
Taylor, M. O'Keeffe, and O. M. Yaghi, Science \textbf{316}, 268 (2007).

    
\bibitem{Han} S. S. Han, H. Furukawa, O. M. Yaghi and W. A. Goddard Iii, J. of the Am. Chem. Soc. \textbf{130} 11580 (2008). 


\bibitem{Mahmood} J. Mahmood et al., 
Nat. Commun. \textbf{6}, 6486 (2015).

\bibitem{H.Sahin} H. Sahin, Phys. Rev. B \textbf{92}, 085421 (2015).

    
\bibitem{Jun} J. Kang, S. Horzum, and F. M. Peeters, Phys. Rev. B \textbf{92} 195419 (2015). 


\bibitem{zhangrev} R. Zhang, B. Li, and J. Yang, Nanoscale \textbf{7}, 14062 (2015).


\bibitem{xu2015} B. Xu, H. Xiang, Q. Wei, J. Q. Liu, Y. D. Xia, J. Yin, Z. G. and Liu, Phys. Chem. Chem. Phys. \textbf{17}, 
15115 (2015). 



\bibitem{vasp1} G. Kresse and J. Hafner, Phys. Rev. B \textbf{47}, 558
(1993).
\bibitem{vasp2} G. Kresse and J. Hafner, Phys. Rev. B \textbf{49}, 14251
(1994).
\bibitem{vasp3} G. Kresse and J. Furthm\"{u}ller, Comput. Mat. Sci. \textbf{6},
15 (1996).
\bibitem{vasp4} G. Kresse and J. Furthm\"{u}ller, Phys. Rev. B \textbf{54}, 
11169
(1996).

\bibitem{Kohn-Sham} W. Kohn and L. J. Sham, Phys. Rev. \textbf{140}, A1133 
(1965).

\bibitem{GGA-PBE1} J. P. Perdew, K. Burke, and M. Ernzerhof, Phys. Rev. Lett.
\textbf{77}, 3865 (1996).
\bibitem{GGA-PBE2} J. P. Perdew, K. Burke, and M. Ernzerhof, Phys. Rev. Lett.
\textbf{78}, 1396 (1997).

\bibitem{HSE} J. Heyd, G. E. Scuseria, and M. Ernzerhof, The J. of Chem. Phys. \textbf{180}, 2622 (2009).

\bibitem{Henkelman} G. Henkelman, A. Arnaldsson, and H. Jonsson,
Comput Mater Sci \textbf{36}, 354 (2006).



\bibitem{CLee} C. Lee, X. Wei, J. W. Kysar, and J. Hone, Science
\textbf{321}, 385 (2008).

\bibitem{Berseneva} N. Berseneva, A. Gulans, A. V. Krasheninnikov, and R. M. Nieminen, 
Phys. Rev. B \textbf{87}, 035404 (2013).





\end{thebibliography}
\end{document}